\documentclass[prb,reprint, amsmath,amssymb,showpacs,superscriptaddress,floatfix]{revtex4-2}
\usepackage[breaklinks=true,colorlinks,citecolor=blue,linkcolor=blue,urlcolor=blue]{hyperref}
\usepackage[table,xcdraw]{xcolor}

\usepackage{epsfig,mathrsfs,color,latexsym,subfigure,marginnote,gensymb,}
\usepackage{graphicx}
\usepackage{xcolor}
\usepackage{booktabs}
\usepackage{appendix}
\renewcommand{\BibitemShut}[1]{}

\begin{document}
\title{Deep Learning Assisted Denoising of Experimental Micrographs}

\author{Owais Ahmad}
\email{owaisah@iitk.ac.in}
\author{Albert Linda}
\author{Saumya Ranjan Jha}
\author{Somnath Bhowmick}
\email{bsomnath@iitk.ac.in}
\affiliation{Department of Materials Science and Engineering, Indian Institute of Technology, Kanpur, Kanpur-208016, UP, India}

\begin{abstract}
Microstructure imaging is crucial in materials science, but experimental images often introduce noise that obscures critical structural details. This study presents a novel deep learning approach for robust microstructure image denoising, combining phase-field simulations, Fourier transform techniques, and an attention-based neural network. The innovative framework addresses dataset limitations by synthetically generating training data by combining computational phase-field microstructures with experimental optical micrographs. The neural network architecture features an attention mechanism that dynamically focuses on important microstructural features while systematically eliminating noise types like scratches and surface imperfections. Testing on a FeMnNi alloy system demonstrated the model's exceptional performance across multiple magnifications. By successfully removing diverse noise patterns while maintaining grain boundary integrity, the research provides a generalizable deep-learning framework for microstructure image enhancement with broad applicability in materials science.
\end{abstract}


\maketitle

\section{Introduction}
Imaging is an important technique used across various scientific and engineering disciplines, like materials science, geology, biomedical applications, etc. In materials science and engineering, high-resolution optical microscopes and electron microscopes are frequently used to capture intricate details of material microstructures~\cite{chaurasia2023novel} at micro and nanometer resolutions. However, these imaging methods often introduce noise and artifacts that can obscure important features and hinder accurate analysis~\cite{panda2019deep}.

The challenge of denoising microstructure images while preserving critical structural information has led to the development of various computational approaches. Traditional methods, including Gaussian filtering and wavelet-based techniques, have shown limited success in handling complex noise patterns without compromising fine details. More recently, various statistical methods and deep learning-based approaches have emerged as powerful tools for image-denoising tasks~\cite{uSCHULZ2023113699,uKUSUMI2024113996}, demonstrating superior performance in preserving edge information and structural integrity. For instance, Zhang et al. proposed a residual learning approach (DnCNN) that achieved state-of-the-art results on synthetic noise datasets \cite{zhang2017beyond}. Another notable contribution is the FFDNet by Zhang et al., which offers a fast and flexible solution for CNN-based image denoising~\cite{zhang2018ffdnet}.

Additionally, the application of Generative Adversarial Networks (GANs) to the specific problem of noise removal in microstructure images has shown promising results~\cite{chen2020image, yang2018low}. Panda et al.~\cite{panda2019deep} addressed refining raw microstructure images of plain carbon steel, focusing on denoising to obtain clean grain surfaces. Their work highlighted several challenges in microstructure image processing, including nonuniform edge width, etching artifacts, and pixel intensity variations due to optical microscopy. They developed a preprocessing framework to generate clean ground-truth images for training their denoising model, addressing the lack of available databases for plain carbon steel microstructure denoising. By posing the denoising task as an image translation problem, their GAN-based approach demonstrated promising results in removing noise while preserving critical microstructural features~\cite{zhu2017unpaired}.

Among deep learning architectures, attention models have gained significant traction due to their ability to focus on relevant features while suppressing noise. Initially popularized in natural language processing tasks, attention mechanisms have been successfully adapted for computer vision applications, including image denoising. The self-attention mechanism, in particular, allows models to capture long-range dependencies in images, making them well-suited for handling the complex patterns found in microstructure data.

\begin{figure*}[ht]
\centering
\includegraphics[width=\textwidth]{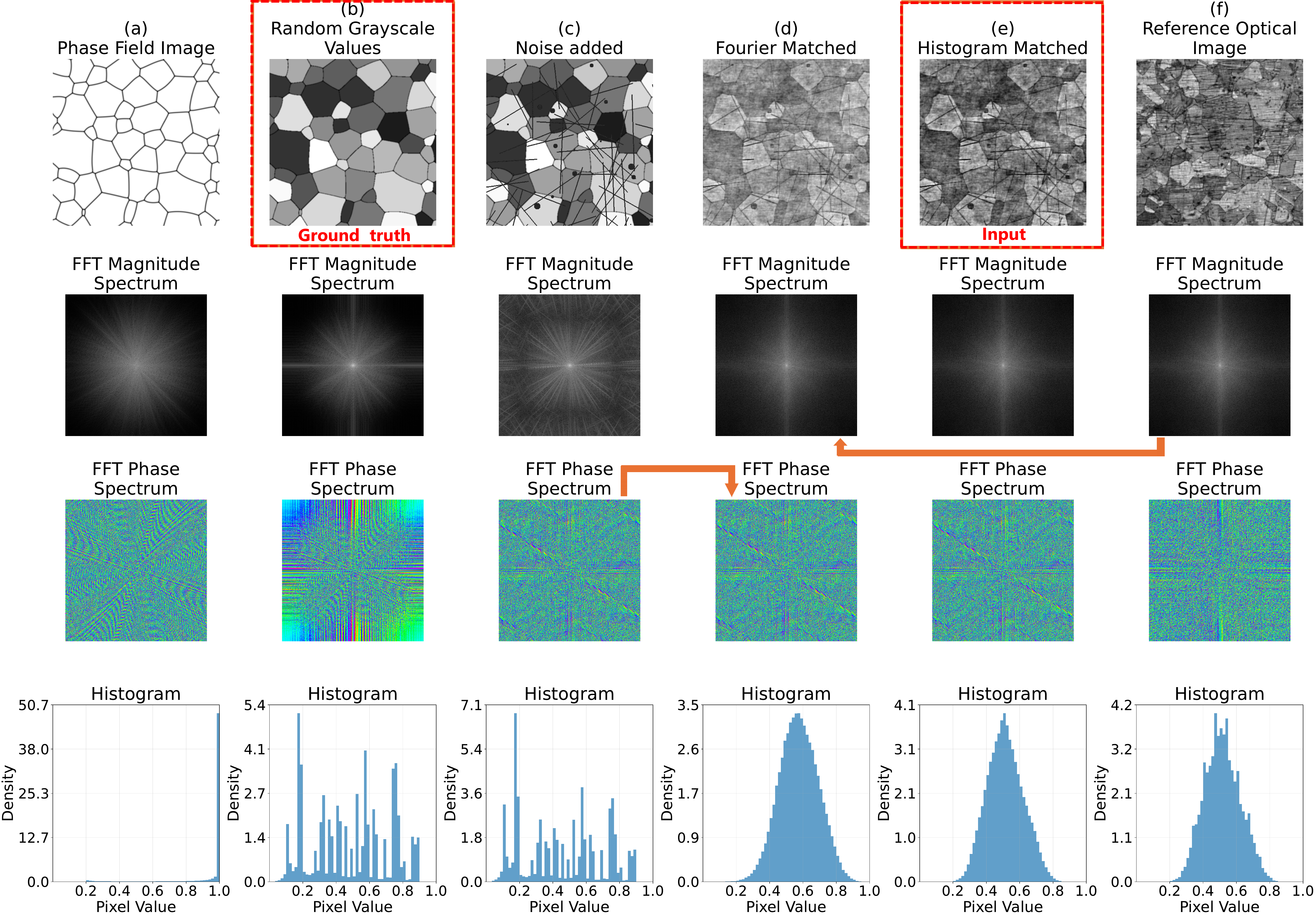}
\caption{Method of combining a phase-field microstructure [panel (a)] and a reference experimental optical micrograph [panel (f)] to create an input image for the training dataset. Image (b) is obtained after adding random grayscale intensities to image (a). Image (c) is constructed by adding artificial noise to image (b). Image (d) is obtained by combining the phase of image (c) and the magnitude of image (f). Panel (e) illustrates the final image after histogram matching for intensity calibration. Image (e) is the input image for training and validation, and image (b) is the corresponding ground truth.}
\label{fig1}
\end{figure*}

Despite the recent advancements, the following challenges remain in applying advanced deep learning models to microstructure denoising.
\begin{itemize}
    \item Adapting attention mechanisms to handle the unique characteristics of different imaging modalities and material types.
    \item Balancing the trade-off between noise reduction and preservation of fine structural details.
    \item Developing efficient training strategies for deep learning models on limited labeled microstructure datasets.
\end{itemize}
This paper addresses these challenges by proposing a novel attention-based architecture designed explicitly for microstructure denoising. We comprehensively evaluate our approach on diverse microstructure datasets and compare its performance against existing state-of-the-art methods. Additionally, we explore the interpretability of our model's attention maps to gain insights into the denoising process and its impact on microstructural feature preservation.

\section{Methodology}
\subsection{Dataset Preparation}
One needs a large number of realistic microstructures to train the model. We adopt a combined computational and experimental approach to build the training set. Before detailing the steps to create the training set, let us briefly discuss the different technical aspects involved in this process.
\paragraph{Simulated and experimental microstructures:}
Generating a realistic microstructure via simulation requires replicating an overall distribution of grains mimicking the size and shape distributions (like grains having curved grain boundaries and different numbers of sides) typically observed in experimental microstructures~\cite{Steinbach_2009, Grnsy2006}. The training set requires many such realistic microstructures, which are obtained using a phase-field model for grain growth. The technical details of the phase-field model are given in Appendix. We also use a finite number of experimental optical micrographs, and details are given in Appendix.

\begin{figure*}[htbp]
\centering
\includegraphics[width=1.8\columnwidth]{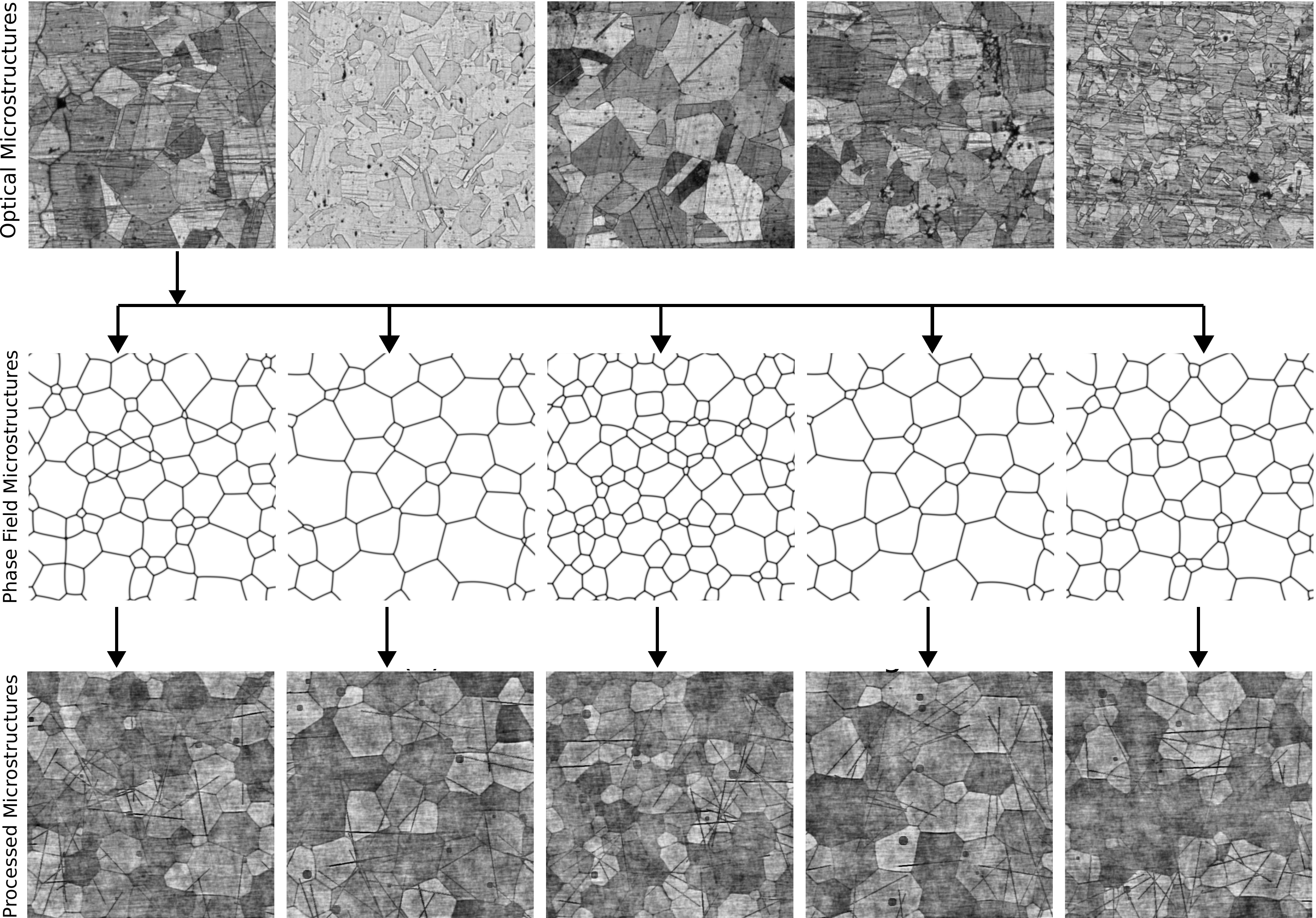}
\caption{Examples of combining phase-field simulated microstructures and experimental optical microstructures to get multiple images for the training dataset.}
\label{fig2}
\end{figure*}

\paragraph{Fast Fourier transform (FFT):} 
FFT breaks down a signal into a sum of sinusoidal components of varying frequencies \cite{brigham1988fast}. In the study of image or microstructure, FFT is commonly applied to analyze and manipulate the frequency components of the data, providing insights into periodic patterns or noise within the spatial domain~\cite{bracewell1986fourier}. During the forward transform, the signal is transformed from the spatial domain (e.g., pixel intensities in an image) to the frequency domain. Each point in the frequency domain has a magnitude and a phase; the former represents the strength of a specific frequency, and the latter encodes the position of the sinusoidal wave relative to a reference point~\cite{davidson2002optical}. In the context of a microstructure, magnitude controls the overall spatial variation of the brightness intensity of the pixels, while the phase determines the spatial arrangement of features (grain shape of a microstructure)~\cite{tsopanidis2021unsupervised,durmaz2021deep}. After the forward transform of a microstructure, both magnitude and phase can be modified in the frequency domain, leading to alteration of the original microstructure. After necessary modification in the frequency domain, the inverse FFT (IFFT) can be applied to transform the modified signal back to the spatial domain.

\paragraph{Combining simulation and experiment:}
The phase-field model generates ``ideal'' microstructures without the noise generally observed in experimental microstructures~\cite{peregrina2013automatic}. Thus, we must add realistic noise from actual optical micrographs to the simulated microstructures to build our training dataset. First, by applying FFT, the magnitude and phase of the frequency components are extracted, both for a simulated micrograph and an actual optical micrograph~\cite{li2020grain, cang2017microstructure}. Next, the phase information from the simulated microstructure is combined with the magnitude of the actual optical microstructure in the Fourier space. This process ensures that we retain the grain shapes from the simulated microstructure while adding some noise and replicating the spatial variation of the brightness intensity of the pixels from the experimental optical microstructure. Finally, an inverse FFT transforms the modified microstructure back to the spatial domain. The final image is enriched with realistic noise and becomes a part of the training dataset of ``noisy images'', while the starting simulated image serves as the ground truth of ``clean images''. Our machine-learning model learns how to map the ``noisy images'' back to the ``clean images''.

Having discussed the technical details, let us describe the step-by-step process of training data generation. We begin by generating the phase-field microstructure [Figure~\ref{fig1}(a)]. Next, we randomly assign different grayscale intensities to the individual grains in the microstructure\cite{banerjee2019automated} [Figure~\ref{fig1}(b)], which becomes the ground truth image. Now, we must add noise to the ground truth images such that they resemble actual experimental micrographs. The first level of noise addition involves introducing random noise in the form of black spots and lines, representing stains and scratches~\cite{campbell2018new}. However, the resulting image [Figure~\ref{fig1}(c)] is far from an actual experimental image. To improve further, we apply the Fourier transform to an experimental optical microstructure [Figure~\ref{fig1}(f)] and extract its magnitude and phase information~\cite{ma2020data}. We perform the same operation for the simulated microstructure [Figure~\ref{fig1}(c)] as well. Then, in the frequency domain, we combine the phase values extracted from the simulated microstructure [Figure~\ref{fig1}(c)] and the magnitude values extracted from the experimental microstructure [Figure~\ref{fig1}(f)]. After performing inverse FFT, we obtain an image like Figure~\ref{fig1}(d), which resembles that of an actual experimental microstructure. Note that Figure~\ref{fig1}(d) retains the grain shapes of the simulated image [Figure~\ref{fig1}(c)]. In addition, it also has defects and spatial variation of the brightness intensity of the pixels similar to experimental optical microstructure [Figure~\ref{fig1}(f)]. We do a final modification by histogram matching between the last image [Figure~\ref{fig1}(d)] and the experimental image [Figure~\ref{fig1}(f)], aligning their intensity distributions. The final image, Figure~\ref{fig1}(e), will be used as the input image for the  ML model, while Figure~\ref{fig1}(b) will serve as the ground truth. Next, we shall train an ML model to map the ``noisy images'' back to the ``clean images'' similar to the ground truth.



Note that since we use an experimental optical image as the reference for adding texture, we can apply the magnitude values from a single experimental optical image to multiple phase-field images. As demonstrated in Figure~\ref{fig2}, this allows us to enhance multiple simulated microstructures using the same reference. For example, suppose we have five phase-field images and five optical microstructures. In that case, we can combine each of the five optical microstructures with each of the five phase-field images, resulting in 25 distinct training images.

\subsection{Model Architecture}
The network is designed to process noisy microstructure images and output clean, denoised versions. Let $I_{noisy}$ and $I_{clean}$ denote the input noisy and output clean images, respectively. One can express the overall network as
\begin{equation}
I_{clean} = f_{reconstruction}(f_{information}(f_{feature}(I_{noisy}))),
\end{equation}
where $f_{feature}$, $f_{information}$, and $f_{reconstruction}$ represent the feature extraction, information extraction, and reconstruction modules respectively. Details of each module are given in the following text.

\begin{figure*}[htbp]
\centering
\includegraphics[width=2\columnwidth]{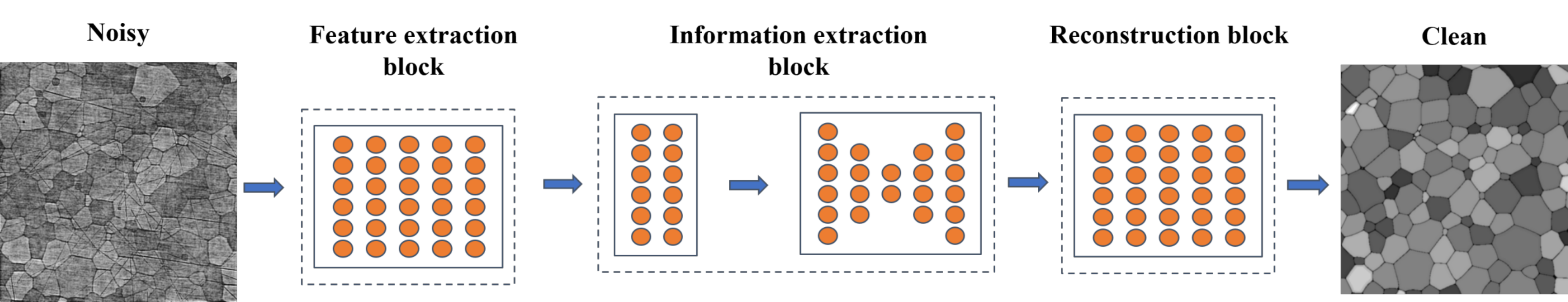}
\caption{Schematic representation of the proposed microstructure denoising model. The architecture consists of three main components: (a) Feature Extraction Block, which captures low-level features from the noisy input; (b) Information Extraction Block, comprising multiple residual blocks with attention mechanisms for noise separation and feature refinement; and (c) Reconstruction Module, which transforms the refined features into a clean microstructure image. Arrows indicate the flow of data through the network.}
\label{fig3}
\end{figure*}

\paragraph{Feature extraction block:} The feature extraction block serves as the initial processing stage for the noisy input image. Its primary role is to capture low-level features and begin the process of separating noise from the underlying microstructure. This block consists of two convolutional layers with batch normalization (BN) and LeakyReLU activation,
\begin{equation}
x_0 = \sigma(BN(Conv_{3\times3}(I_{noisy}))),
\end{equation}
\begin{equation}
x_1 = \sigma(BN(Conv_{3\times3}(x_0))),
\end{equation}
where $Conv_{3\times3}$ denotes a $3\times 3$ convolutional layer, $BN$ is batch normalization, and $\sigma$ represents the LeakyReLU activation function with $\alpha=0.2$. The first convolutional layer aims to detect edges, textures, and other basic structures in the noisy image. The second layer begins to compose these features into more complex patterns. The LeakyReLU activation introduces non-linearity, allowing the network to learn more sophisticated representations, while batch normalization helps stabilize the learning process and reduces internal covariate shift, which is particularly important when dealing with varying noise levels in microstructure images. 

\paragraph{Information extraction block:} The information extraction block is the core of our denoising process. It combines residual learning and attention mechanisms to effectively separate noise from the true microstructure features. This block consists of multiple residual blocks with attention mechanisms. Each residual block can be expressed as
\begin{equation}
x_{res} = x + \sigma(BN(Conv_{3\times3}(\sigma(BN(Conv_{3\times3}(x)))))).
\end{equation}
The residual structure allows the network to learn the difference between the noisy input and the clean output, which is often easier than directly learning the clean image. This is particularly effective for denoising, as noise can be considered a residual on top of the clean image. The attention mechanism is applied after each residual block,
\begin{equation}
y = \sigma(W_2(\sigma(W_1(GAP(x_{res}))))),
\end{equation}
\begin{equation}
x_{att} = x_{res} \odot y,
\end{equation}
where $GAP$ is global average pooling, $W_1$ and $W_2$ are dense layers, and $\odot$ denotes element-wise multiplication. The attention mechanism allows the network to focus on important features of the microstructure while suppressing noise. It does this by generating a set of weights (y) that are applied to the feature maps. This is crucial for preserving fine details of the microstructure while removing noise, as it allows the network to decide which features are important to retain adaptively. The information extraction block is repeated multiple times, with a dropout layer after each block,
\begin{equation}
x_{n+1} = Dropout(f_{information}(x_n)).
\end{equation}
This repetition allows for progressive refinement of the features, gradually removing more noise at each stage. The dropout layers help prevent overfitting and improve the network's generalization, which is vital for handling various microstructures and noise patterns \cite{kingma2014adam}. 

\paragraph{Reconstruction block:}
The reconstruction block transforms the extracted and refined features into a clean image~\cite{son2017retinal}. It consists of two convolutional layers with batch normalization and LeakyReLU activation, followed by a final convolutional layer with sigmoid activation
\begin{equation}
x_{rec1} = \sigma(BN(Conv_{3\times3}(x_n))),
\end{equation}
\begin{equation}
x_{rec2} = \sigma(BN(Conv_{3\times3}(x_{rec1}))),
\end{equation}
\begin{equation}
I_{clean} = sigmoid(Conv_{3\times3}(x_{rec2})).
\end{equation}

The first two layers in this module further refine the features, potentially removing any remaining noise artifacts~\cite{radford2016unsupervised}. The final layer with sigmoid activation ensures the output is appropriately scaled to represent a clean image. This module plays a crucial role in ensuring that the final output retains the critical structural details of the microstructure while being free of the original noise \cite{chen2018efficient}.

\paragraph{Loss function and training:}
The network is trained using a combination of mean squared error (MSE) loss and L2 regularization
\begin{equation}
\mathcal{L} = MSE(I_{clean}, I_{gt}) + \lambda \sum_i ||w_i||^2_2,
\end{equation}
where $I_{gt}$ is the ground truth clean image, $w_i$ are the weights of the network, and $\lambda$ is the regularization coefficient. The MSE loss encourages the network to produce outputs that are pixel-wise similar to the ground truth clean images. This method is effective for denoising as it directly penalizes differences between the predicted clean image and the actual clean image. The L2 regularization term helps prevent overfitting by encouraging smaller weight values, which is important for the network's ability to generalize to unseen microstructure images. The network is optimized using the Adam optimizer with a learning rate of $10^{-4}$ and default beta values. This adaptive optimization method helps navigate the complex loss landscape associated with denoising tasks, allowing for efficient training even with the intricate patterns present in microstructure images.

\section{Results and Discussion}
\begin{figure*}[htbp]
\centering
\includegraphics[width=2\columnwidth]{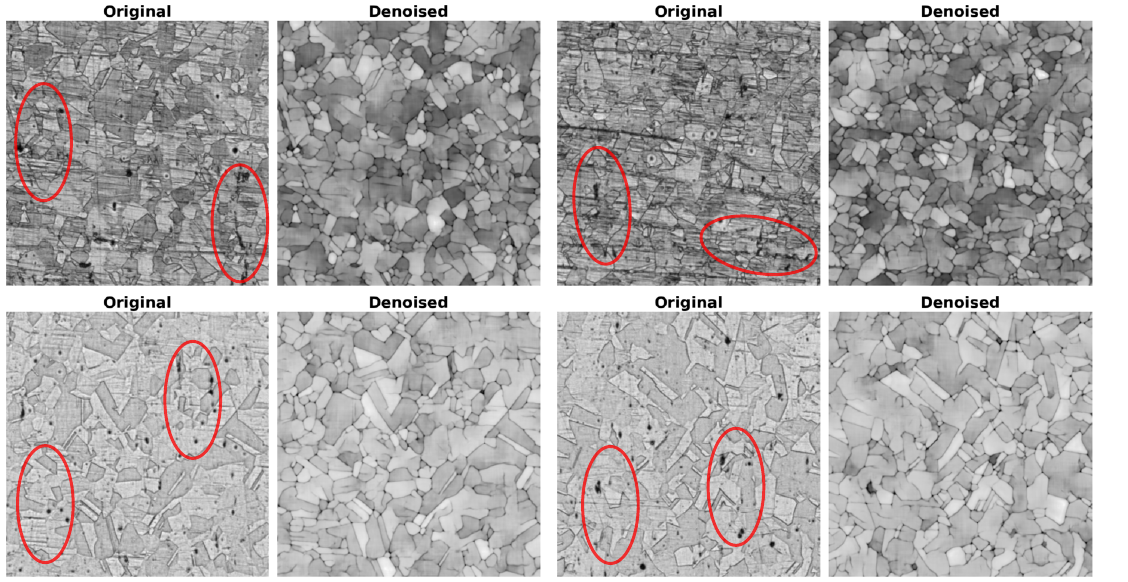}
\caption{Comparison of original and denoised microstructure images: Original optical microstructures contain various types of noise, including scratches, pit stains, and minor surface imperfections. Corresponding denoised images produced by our model demonstrate effective noise removal while preserving grain boundaries and overall microstructure integrity.}
\label{fig4}
\end{figure*}

Using the method illustrated in Figure~\ref{fig1} and Figure~\ref{fig2}, we construct a dataset comprising 3,000 images, which are partitioned into training and testing sets using an 80:20 split. Specifically, 2,400 images are allocated for training, with the remaining 600 images reserved for model validation. The training is conducted on an Intel\textregistered{} Xeon\textregistered{} Gold 6338N CPU (2.20~GHz), with the computational process requiring approximately 6 hours. Subsequently, the trained model demonstrates prediction times averaging approximately 5 seconds per inference.

The performance of our trained model demonstrates significant improvements in the quality of microstructure images, as illustrated in Figure~\ref{fig4}. The model successfully addresses a broad spectrum of noise types commonly encountered in optical microstructures, ranging from deep scratches to pit stains and numerous minor surface imperfections. The following text elaborates on the key features of the model.

\paragraph{Noise removal and grain boundary detection:}
One of the most notable achievements of our model is its ability to differentiate between various forms of image artifacts and genuine microstructural features like grain boundaries. The latter is crucial for maintaining the structural integrity of the microstructure representation. As Figure~\ref{fig4} highlights, the model effectively removes different noises like polishing scratches (deep as well as faint), water stains, etch pits, speckle patterns, and corroded thick boundary layers. It is pertinent to note that even faint polishing marks can be efficiently removed to improve granular contrast, irrespective of the magnification and resolution of the image. This comprehensive noise removal is achieved while preserving the integrity of the grain boundaries, which is a critical aspect in microstructure analyses. The model's capability to distinguish between noise, artifacts, and essential microstructural features highlights its ability to learn to recognize the fundamental characteristics that differentiate grain boundaries from artifacts, even when they may appear visually similar. Aside from the visual aesthetics, this model renders the microstructure suitable for automated grain size and distribution-based analyses. Furthermore, its capability to discern not just polygonal grain boundaries but also even effectively differentiate twin boundaries underlines its robustness and versatility.


\paragraph{Intensity adjustment and histogram matching:}
Our analysis reveals that denoising slightly alters the individual grain intensities in the predicted images. We implement a post-processing step using histogram matching to address this and ensure fidelity to the original image characteristics. This technique adjusts the intensity distribution of the denoised image to match that of the original image. The histogram equalization process is particularly effective because scratch values typically constitute only a tiny portion of the overall pixel intensities. As a result, the final denoised image maintains pixel values in the grain regions that are very close to those in the original image, preserving the overall appearance and intensity distribution of the microstructure.
The diverse range of noise types successfully removed by our model suggests a high degree of generalization. This process is particularly important in real-world scenarios where optical microstructures may be affected by various environmental factors and imaging conditions. The model's robustness in handling different noise patterns indicates its potential applicability across a wide range of materials and imaging setups.

Retention of granular contrast (without explicit training) asserts that the methodology may be extended to orientation-based imaging techniques such as electron channeling contrast imaging (ECCI) and electron backscatter diffraction (EBSD) images as well for a plethora of microstructural analyses. Noise removal facilitates accurate grain size analysis, and our tool can be extended to incorporate that capability. One could extend it to grain boundary character analyses with the help of carefully curated labeled data.

\section{Conclusion}
In conclusion, we demonstrate a deep learning-based method to denoise microstructure images while preserving crucial structural information. The combination of effective noise removal, accurate grain boundary detection, and intensity preservation through histogram matching results in high-quality, realistic microstructure images. These enhanced images can significantly improve the accuracy and reliability of microstructure analysis in materials science and engineering applications.

\appendix
\label{pfm}
\section{Phase field model}
The microstructure is described by a set of non-conserved order parameters $\eta_1$, $\eta_2$,...,$\eta_Q$, each representing a type of grain. The evolution of these order parameters over time is governed by the Allen-Cahn equation~\cite{ALLEN19791085}, which describes the kinetics of interface motion. The Allen-Cahn equation is expressed as
\begin{equation}
   \frac{\partial \eta_{i}}{\partial t} = -L_{i}\frac{\delta F}{\delta \eta_{i}}, i = 1,2...Q.
\end{equation}
Here $L_i$ is the kinetic coefficient, and  $F$ is the free energy functional, which is typically written as a sum of two contributions. A local free energy density $f(\eta_1,\eta_2...\eta_Q)$ captures the thermodynamic interactions between different grains, and a gradient energy term penalizes spatial variations in the order parameters to account for interfacial energy. The free energy functional is expressed as
\begin{equation}
   F = \left[f( \eta_{1}, \eta_{2},...\eta_{Q}) + \sum_{i=1}^{Q}\frac{\kappa_{i}}{2}|\nabla \eta_{i}|^2\right]dV.
\end{equation}
In the above expression, $\kappa_i$ represents the gradient energy coefficient, which controls the width of the interfaces between grains. The gradient energy term $\frac{\kappa_i}{2}|\nabla \eta_i|^2$ ensures that sharp interfaces are smoothed over a finite thickness. The local free energy term $f$ determines the stability of individual grains. In this work, $f$ is chosen to have the following form,
\begin{equation}
   f(\eta_{1}, \eta_{2},...\eta_{Q}) = \sum_{i=1}^{Q}(-\frac{1}{2}\alpha \eta_{i}^2 + \frac{1}{4} \beta \eta_{i}^4) + \gamma \sum_{i=1}^{Q} \sum_{j\neq i}^{Q}\eta_{i}^{2}\eta_{j}^{2}.
\end{equation}
In the above expression, $\alpha$ and $\beta$ determine the double-well potential. The second term introduces a coupling between the order parameters, which discourages overlap between grains by penalizing regions where multiple $\eta_i$'s are non-zero. Combining all the terms, one can write the governing equation for the temporal evolution of the order parameters as,
\begin{equation}
   \frac{\partial \eta_{i}}{\partial t} = -L_{i} \left( -\alpha\eta_{i} + \beta\eta_{i}^3 + 2\gamma\eta_{i} \sum_{j\neq i}^{Q} \eta_{j}^2 -\kappa_{i}\nabla^2\eta_{i} \right).
   \label{eq:allan_cahn}
\end{equation}
The initial microstructure is generated using the Voronoi tessellation with almost equal grain size throughout the region. Then, we numerically solve  Equation~\ref{eq:allan_cahn} using the finite difference method.

\section{Optical microscopy}
A single-phase FCC alloy system of equiatomic FeMnNi was recrystallized at 1000$^{\circ}$C for 2 hours, followed by ice brine quenching to retain the single-phase FCC behavior. This method assures that any granular contrast achieved post-etching would be orientation-based due to the differential absorption of grains depending on their free energies and not because of any phase contrast. 10${\%}$  Nital reagent is used to etch the metallographically polished specimens for 20 s, 30 s, and 40 s time intervals. Optical microscopy images are acquired utilizing a Leica DM6000-M optical microscope (Leica Microsystems CMS GmbH, Switzerland) at magnifications of 100X, 200X, and 500X for each of the specimens, using the Leica application Suite v3.4.1 software.

\section*{Acknowledgements}
We acknowledge the National Super Computing Mission (NSM) for providing computing resources of “PARAM Sanganak” at IIT Kanpur, which is implemented by CDAC and supported by the Ministry of Electronics and Information Technology (MeitY) and Department of Science and Technology (DST), Government of India. We also thank ICME National Hub, IIT Kanpur and CC, IIT Kanpur for providing the HPC facility.


\begin{thebibliography}{10}
\expandafter\ifx\csname url\endcsname\relax
  \def\url#1{\texttt{#1}}\fi
\expandafter\ifx\csname urlprefix\endcsname\relax\def\urlprefix{URL }\fi
\expandafter\ifx\csname href\endcsname\relax
  \def\href#1#2{#2} \def\path#1{#1}\fi

\bibitem{chaurasia2023novel}
N.~Chaurasia, S.~K. Jha, S.~Sangal,
  \href{http://dx.doi.org/10.1557/jmr.2006.0011}{A novel training methodology
  for phase segmentation of steel microstructures using a deep learning
  algorithm}, Materialia 30 (2023) 101803.
\newblock \href {https://doi.org/10.1016/j.mtla.2023.101803}
  {\path{doi:10.1016/j.mtla.2023.101803}}.
\newline\urlprefix\url{http://dx.doi.org/10.1557/jmr.2006.0011}

\bibitem{panda2019deep}
A.~Panda, R.~Naskar, S.~Pal,
  \href{https://doi.org/10.1049/iet-ipr.2018.6302}{Deep learning approach for
  segmentation of plain carbon steel microstructure images}, IET Image
  Processing 13~(9) (2019) 1516--1524.
\newblock \href {https://doi.org/10.1049/iet-ipr.2018.6302}
  {\path{doi:10.1049/iet-ipr.2018.6302}}.
\newline\urlprefix\url{https://doi.org/10.1049/iet-ipr.2018.6302}

\bibitem{uSCHULZ2023113699}
B.~Schulz, N.~Haghdadi, T.~Leitner, M.~Hafok, S.~Primig,
  \href{https://www.sciencedirect.com/science/article/pii/S0304399123000165}{Advancing
  analytical electron microscopy methodologies to characterise microstructural
  features in superalloys}, Ultramicroscopy 247 (2023) 113699.
\newblock \href
  {https://doi.org/https://doi.org/10.1016/j.ultramic.2023.113699}
  {\path{doi:https://doi.org/10.1016/j.ultramic.2023.113699}}.
\newline\urlprefix\url{https://www.sciencedirect.com/science/article/pii/S0304399123000165}

\bibitem{uKUSUMI2024113996}
T.~Kusumi, S.~Katakami, R.~Ishikawa, K.~Kawahara, T.~Mullarkey, J.~M.
  Bekkevold, J.~J. Peters, L.~Jones, N.~Shibata, M.~Okada,
  \href{https://www.sciencedirect.com/science/article/pii/S0304399124000755}{New
  poisson denoising method for pulse-count stem imaging}, Ultramicroscopy 264
  (2024) 113996.
\newblock \href
  {https://doi.org/https://doi.org/10.1016/j.ultramic.2024.113996}
  {\path{doi:https://doi.org/10.1016/j.ultramic.2024.113996}}.
\newline\urlprefix\url{https://www.sciencedirect.com/science/article/pii/S0304399124000755}

\bibitem{zhang2017beyond}
K.~Zhang, W.~Zuo, Y.~Chen, D.~Meng, L.~Zhang,
  \href{https://doi.org/10.1109/TIP.2017.2662206}{Beyond a {G}aussian denoiser:
  Residual learning of deep {CNN} for image denoising}, IEEE Transactions on
  Image Processing 26~(7) (2017) 3142--3155.
\newblock \href {https://doi.org/10.1109/TIP.2017.2662206}
  {\path{doi:10.1109/TIP.2017.2662206}}.
\newline\urlprefix\url{https://doi.org/10.1109/TIP.2017.2662206}

\bibitem{zhang2018ffdnet}
K.~Zhang, W.~Zuo, L.~Zhang,
  \href{https://doi.org/10.1109/TIP.2018.2839891}{{FFDNet}: Toward a fast and
  flexible solution for {CNN}-based image denoising}, IEEE Transactions on
  Image Processing 27~(9) (2018) 4608--4622.
\newblock \href {https://doi.org/10.1109/TIP.2018.2839891}
  {\path{doi:10.1109/TIP.2018.2839891}}.
\newline\urlprefix\url{https://doi.org/10.1109/TIP.2018.2839891}

\bibitem{chen2020image}
S.~Chen, D.~Shi, M.~Sadiq, X.~Cheng,
  \href{https://doi.org/10.1109/access.2020.2988284}{Image denoising with
  generative adversarial networks and its application to cell image
  enhancement}, IEEE Access 8 (2020) 82819--82831.
\newblock \href {https://doi.org/10.1109/access.2020.2988284}
  {\path{doi:10.1109/access.2020.2988284}}.
\newline\urlprefix\url{https://doi.org/10.1109/access.2020.2988284}

\bibitem{yang2018low}
Q.~Yang, P.~Yan, Y.~Zhang, H.~Yu, Y.~Shi, X.~Mou, M.~K. Kalra, G.~W. Zhang,
  \href{https://doi.org/10.1109/TMI.2018.2827462}{Low-dose {CT} image denoising
  using a generative adversarial network with {W}asserstein distance and
  perceptual loss}, IEEE Transactions on Medical Imaging 37~(6) (2018)
  1348--1357.
\newblock \href {https://doi.org/10.1109/TMI.2018.2827462}
  {\path{doi:10.1109/TMI.2018.2827462}}.
\newline\urlprefix\url{https://doi.org/10.1109/TMI.2018.2827462}

\bibitem{zhu2017unpaired}
J.-Y. Zhu, T.~Park, P.~Isola, A.~A. Efros,
  \href{https://doi.org/10.1109/ICCV.2017.244}{Unpaired image-to-image
  translation using cycle-consistent adversarial networks}, Proceedings of the
  IEEE International Conference on Computer Vision (2017) 2223--2232\href
  {https://doi.org/10.1109/ICCV.2017.244} {\path{doi:10.1109/ICCV.2017.244}}.
\newline\urlprefix\url{https://doi.org/10.1109/ICCV.2017.244}

\bibitem{Steinbach_2009}
I.~Steinbach,
  \href{https://dx.doi.org/10.1088/0965-0393/17/7/073001}{Phase-field models in
  materials science}, Modelling and Simulation in Materials Science and
  Engineering 17~(7) (2009) 073001.
\newblock \href {https://doi.org/10.1088/0965-0393/17/7/073001}
  {\path{doi:10.1088/0965-0393/17/7/073001}}.
\newline\urlprefix\url{https://dx.doi.org/10.1088/0965-0393/17/7/073001}

\bibitem{Grnsy2006}
L.~Gr$\acute{a}$n$\acute{a}$sy, T.~Pusztai, T.~B\"{o}rzs\"{o}nyi,
  G.~T$\acute{o}$th, G.~Tegze, J.~Warren, J.~Douglas,
  \href{http://dx.doi.org/10.1557/jmr.2006.0011}{Phase field theory of crystal
  nucleation and polycrystalline growth: A review}, Journal of Materials
  Research 21~(2) (2006) 309--319.
\newblock \href {https://doi.org/10.1557/jmr.2006.0011}
  {\path{doi:10.1557/jmr.2006.0011}}.
\newline\urlprefix\url{http://dx.doi.org/10.1557/jmr.2006.0011}

\bibitem{brigham1988fast}
E.~O. Brigham, The Fast {F}ourier Transform and its Applications, Prentice
  Hall, 1988.

\bibitem{bracewell1986fourier}
R.~N. Bracewell, The {F}ourier Transform and its Applications, Vol. 31999,
  McGraw-Hill, 1986.

\bibitem{davidson2002optical}
M.~W. Davidson, M.~Abramowitz,
  \href{https://doi.org/10.1002/0471443395.img074}{Optical microscopy},
  Encyclopedia of Imaging Science and Technology 2 (2002) 1106--1141.
\newblock \href {https://doi.org/10.1002/0471443395.img074}
  {\path{doi:10.1002/0471443395.img074}}.
\newline\urlprefix\url{https://doi.org/10.1002/0471443395.img074}

\bibitem{tsopanidis2021unsupervised}
S.~Tsopanidis, S.~Osovski,
  \href{https://doi.org/10.1016/j.matchar.2021.111551}{Unsupervised machine
  learning in fractography: Evaluation and interpretation}, Materials
  Characterization 182 (2021) 111551.
\newblock \href {https://doi.org/10.1016/j.matchar.2021.111551}
  {\path{doi:10.1016/j.matchar.2021.111551}}.
\newline\urlprefix\url{https://doi.org/10.1016/j.matchar.2021.111551}

\bibitem{durmaz2021deep}
A.~R. Durmaz, M.~Müller, B.~Lei, A.~Thomas, D.~Britz, E.~A. Holm, C.~Eberl,
  F.~Mücklich, P.~Gumbsch, \href{https://doi.org/10.1038/s41467-021-26565-5}{A
  deep learning approach for complex microstructure inference}, Nature
  Communications 12~(1) (2021) 6272.
\newblock \href {https://doi.org/10.1038/s41467-021-26565-5}
  {\path{doi:10.1038/s41467-021-26565-5}}.
\newline\urlprefix\url{https://doi.org/10.1038/s41467-021-26565-5}

\bibitem{peregrina2013automatic}
H.~Peregrina-Barreto, I.~Terol-Villalobos, J.~Rangel-Magdaleno,
  A.~Herrera-Navarro, L.~Morales-Hernández, F.~Manríquez-Guerrero,
  \href{https://doi.org/10.1016/j.measurement.2012.06.012}{Automatic grain size
  determination in microstructures using image processing}, Measurement 46~(1)
  (2013) 249--258.
\newblock \href {https://doi.org/10.1016/j.measurement.2012.06.012}
  {\path{doi:10.1016/j.measurement.2012.06.012}}.
\newline\urlprefix\url{https://doi.org/10.1016/j.measurement.2012.06.012}

\bibitem{li2020grain}
M.~Li, D.~Chen, S.~Liu,
  \href{https://doi.org/10.1109/access.2020.3011703}{Grain boundary detection
  based on multi-level loss from feature and adversarial learning}, IEEE Access
  8 (2020) 135640--135651.
\newblock \href {https://doi.org/10.1109/access.2020.3011703}
  {\path{doi:10.1109/access.2020.3011703}}.
\newline\urlprefix\url{https://doi.org/10.1109/access.2020.3011703}

\bibitem{cang2017microstructure}
R.~Cang, Y.~Xu, S.~Chen, Y.~Liu, Y.~Jiao, M.~Y. Ren,
  \href{https://doi.org/10.1115/1.4036649}{Microstructure representation and
  reconstruction of heterogeneous materials via deep belief network for
  computational material design}, Journal of Mechanical Design 139~(7) (2017)
  071404.
\newblock \href {https://doi.org/10.1115/1.4036649}
  {\path{doi:10.1115/1.4036649}}.
\newline\urlprefix\url{https://doi.org/10.1115/1.4036649}

\bibitem{banerjee2019automated}
S.~Banerjee, P.~C. Chakraborti, S.~K. Saha,
  \href{https://doi.org/10.1016/j.measurement.2019.03.046}{An automated
  methodology for grain segmentation and grain size measurement from optical
  micrographs}, Measurement 140 (2019) 142--150.
\newblock \href {https://doi.org/10.1016/j.measurement.2019.03.046}
  {\path{doi:10.1016/j.measurement.2019.03.046}}.
\newline\urlprefix\url{https://doi.org/10.1016/j.measurement.2019.03.046}

\bibitem{campbell2018new}
A.~Campbell, P.~Murray, E.~Yakushina, S.~Marshall, W.~Ion,
  \href{https://doi.org/10.1016/j.matdes.2017.12.049}{New methods for automatic
  quantification of microstructural features using digital image processing},
  Materials \& Design 141 (2018) 395--406.
\newblock \href {https://doi.org/10.1016/j.matdes.2017.12.049}
  {\path{doi:10.1016/j.matdes.2017.12.049}}.
\newline\urlprefix\url{https://doi.org/10.1016/j.matdes.2017.12.049}

\bibitem{ma2020data}
B.~Ma, X.~Wei, C.~Liu, X.~Ban, H.~Huang, H.~Wang, W.~Xue, S.~Wu, M.~Gao,
  Q.~Shen, et~al., \href{https://doi.org/10.1038/s41524-020-00392-6}{Data
  augmentation in microscopic images for material data mining}, npj
  Computational Materials 6~(1) (2020) 1--9.
\newblock \href {https://doi.org/10.1038/s41524-020-00392-6}
  {\path{doi:10.1038/s41524-020-00392-6}}.
\newline\urlprefix\url{https://doi.org/10.1038/s41524-020-00392-6}

\bibitem{kingma2014adam}
D.~P. Kingma, J.~Ba, \href{https://arxiv.org/abs/1412.6980}{Adam: A method for
  stochastic optimization}, arXiv preprint arXiv:1412.6980 (2014).
\newline\urlprefix\url{https://arxiv.org/abs/1412.6980}

\bibitem{son2017retinal}
J.~Son, S.~J. Park, K.-H. Jung,
  \href{https://doi.org/10.1007/s10278-018-0126-3}{Retinal vessel segmentation
  in fundoscopic images with generative adversarial networks}, Journal of
  Digital Imaging 32 (2017) 499--512.
\newblock \href {https://doi.org/10.1007/s10278-018-0126-3}
  {\path{doi:10.1007/s10278-018-0126-3}}.
\newline\urlprefix\url{https://doi.org/10.1007/s10278-018-0126-3}

\bibitem{radford2016unsupervised}
A.~Radford, L.~Metz, S.~Chintala,
  \href{https://arxiv.org/abs/1511.06434}{Unsupervised representation learning
  with deep convolutional generative adversarial networks}, Proceedings of 4th
  International Conference on Learning Representations (2016).
\newline\urlprefix\url{https://arxiv.org/abs/1511.06434}

\bibitem{chen2018efficient}
Y.~Chen, A.~G. Christodoulou, Z.~Zhou, F.~Shi, Y.~Xie, D.~Li,
  \href{https://doi.org/10.1007/978-3-030-00928-1_11}{Efficient and accurate
  {MRI} super-resolution using a generative adversarial network and {3D}
  multilevel densely connected network}, Proceedings of International
  Conference on Medical Image Computing and Computer-Assisted Intervention
  (2018) 91--99\href {https://doi.org/10.1007/978-3-030-00928-1_11}
  {\path{doi:10.1007/978-3-030-00928-1_11}}.
\newline\urlprefix\url{https://doi.org/10.1007/978-3-030-00928-1_11}

\bibitem{ALLEN19791085}
S.~M. Allen, J.~W. Cahn,
  \href{https://www.sciencedirect.com/science/article/pii/0001616079901962}{A
  microscopic theory for antiphase boundary motion and its application to
  antiphase domain coarsening}, Acta Metallurgica 27~(6) (1979) 1085--1095.
\newblock \href {https://doi.org/https://doi.org/10.1016/0001-6160(79)90196-2}
  {\path{doi:https://doi.org/10.1016/0001-6160(79)90196-2}}.
\newline\urlprefix\url{https://www.sciencedirect.com/science/article/pii/0001616079901962}

\end{thebibliography}

\end{document}